\documentclass[]{interact}

\usepackage{lineno,hyperref}

\usepackage[numbers,sort&compress]{natbib}
\bibpunct[, ]{[}{]}{,}{n}{,}{,}
\renewcommand\bibfont{\fontsize{10}{12}\selectfont}
\makeatletter
\def\NAT@def@citea{\def\@citea{\NAT@separator}}
\makeatother

\theoremstyle{plain}

\theoremstyle{definition}

\theoremstyle{remark}

\usepackage{graphicx}
\usepackage{bm}
\usepackage{dcolumn}
\usepackage{setspace}

\usepackage{amsmath}
\usepackage{fancyhdr}

\usepackage{caption}
\usepackage{subcaption}

\usepackage{color,soul}

\begin{document}

\articletype{Article}

\title{Size effects on dislocation starvation in Cu nanopillars : A molecular dynamics simulations study}

\author{
\name{G. Sainath\textsuperscript{a,b}\thanks{CONTACT G. Sainath. Email: sg@igcar.gov.in}, Vani Shankar\textsuperscript{a,b} and A. Nagesha\textsuperscript{a,b}}
\affil{\textsuperscript{a}Mechanical Metallurgy Division, Metallurgy and Materials Group, Indira Gandhi Centre for Atomic Research, Kalpakkam, Tamilnadu-603102, India\\
\textsuperscript{b}Homi Bhabha National Institute, Training School Complex, Anushaktinagar, Mumbai, Maharashtra-400094, India}
}

\maketitle

\begin{abstract}
Size plays an important role on the deformation mechanism of nanopillars. With decreasing size, many FCC nanopillars exhibit dislocation starvation state which 
is responsible for their high strength. However, many details about the dislocation starvation mechanism like how often it occurs, and how much is its contribution 
to the total plastic strain, are still elusive. Similarly, the size below which the dislocation starvation occurs in the nanopillars is not clearly established. In 
this context, the atomistic simulations have been performed on the compressive deformation of $<$110$>$ Cu nanopillars with size (d) ranging from 5 to 21.5 nm. The 
molecular dynamics (MD) simulation results indicate that the nanopillars deform by the slip of extended dislocations and exhibit dislocation starvation mainly at 
small sizes ( $<$ 20 nm). The frequency of the occurrence of dislocation starvation is highest in small size nanowires and it decreases with increasing size. Above 
the nanopillar size of 20 nm, no dislocation starvation has been observed. Further, we define the dislocation starvation strain and based on this, it has been shown 
that, the contribution of dislocation starvation state to the total plastic strain decreases from 70\% in small size nanopillars to below 5\% in large size pillars. 
The present results suggest that the dislocation starvation is a dominant phenomena in small size nanopillars.
\end{abstract}


\section{Introduction}
It is well known that the mechanical properties and associated plastic deformation behaviour of crystalline materials is mainly governed by the movement of 
partial or full dislocations and depends mainly on temperature and strain rate. In bulk materials, the yield stress and the plastic deformation mechanisms 
were independent of size. However, when the sample size is reduced to nanoscale, the mechanical properties along with plastic deformation mechanisms exhibit 
strong size dependence \cite{Uchic,Oh}. The high surface area to volume ratio and low density of defects provides the nanostructures with mechanical properties 
different from bulk single crystals. The plastic deformation in nanowires/nanopillars occurs either by dislocation slip or twinning depending on the 
crystallographic orientation \cite{Park-JMPS,Weinberger-Cai}. Many studies over the last two decades have suggested that the size also influence the operative 
deformation mechanisms \cite{Uchic,Oh,Nanolett-1,Ag-sizeffects}. Using in-situ transmission electron microscopy (TEM) experiments on Ag nanowires, a transition 
in deformation mechanisms has been reported with increasing size \cite{Ag-sizeffects}. In Ag nanowires of 11 nm diameter, the deformation proceeds by the slip 
of full dislocations, while nucleation and annihilation of stacking faults has been observed in relatively smaller diameters in the range 5-8 nm. In sub 3 nm 
size nanowires, the deformation is accommodated by the relative slip between two adjacent glide planes without any dislocations \cite{Ag-sizeffects}. Similarly, 
in-situ TEM experiments on gold revealed that thick films deformed predominantly by perfect dislocations, while thin films deformed mainly by partial dislocations 
separated by stacking faults \cite{Oh}. Like deformation mechanism, the yield strength of FCC nanopillars also shows strong size dependence. The yield strength of 
nanowires/nanopillars increases with decreasing diameter and it is significantly higher than their bulk counterparts \cite{Uchic,Nanolett-1}. In order to explain 
this 'smaller is stronger' trend in nanopillars, the dislocation starvation has been proposed as one of the strengthening mechanisms in small size nanopillars 
\cite{Greer-Nix}. The mechanism of dislocation starvation involves the continuous escape of dislocations to nearby free surfaces without any multiplication 
leading to a scarcity of dislocations in the nanopillar. This dislocation free state of the nanopillar during the plastic deformation is known as dislocation 
starvation state \cite{Greer-Nix} or mechanical annealing \cite{Shan-annealing}. This mechanism originates from the micro-structural parameters and dimensional 
constraints such as high stress and small size \cite{Kiritani,Greer-TSF}, which is readily satisfied in nanowires and nanofilms. Following dislocation starvation, 
the nanopillar deforms elastically until new dislocations nucleate to continue further plastic deformation. This continuous nucleation of dislocations requires 
high stress which leads to dislocation starvation hardening \cite{Greer-TSF}. Experimentally, the dislocation starvation mechanism has been observed in many 
FCC nanowires, nanopillars and nanofilms \cite{Greer-Nix,Shan-annealing,Greer-TSF,nanocubes}.\\

Investigating the dislocation starvation related mechanism using conventional experimental methods is quite tedious and requires sophisticated instruments. 
In this context, computational methods based on molecular dynamics (MD) and dislocation dynamics (DD) simulations are adopted to understand the dislocation 
related behaviours in nanopillars \cite{Greer-Cai-MSEA,Cai-PNAS,ACao,DD-starve}. The 3D DD simulation study by Zhou et al. \cite{DD-starve} on Ni micropillars 
has revealed that the dislocation starvation at an early stage of plastic deformation was due to weakly entangled dislocations leaving the sample. The DD study 
by Greer et al. \cite{Greer-Cai-MSEA} and MD simulations study by Weinberger and Cai \cite{Cai-PNAS} have demonstrated that the dislocation starvation occurs 
only in FCC nanopillars and no dislocation starvation has been observed in BCC nanopillars. MD simulations study by Cao et al. \cite{ACao} on Ni nanopillars 
of diameters ranging from 4 to 16 nm has revealed that in dislocation starvation state, the nanopillars deform elastically until new dislocations nucleate 
from the surfaces. Despite these experimental and simulation studies \cite{Greer-Nix,Shan-annealing,Greer-TSF,nanocubes,Greer-Cai-MSEA,Cai-PNAS,ACao,DD-starve}, 
many details about the dislocation starvation are still elusive. For example, how often the dislocation starvation occurs during the plastic deformation and 
what is the contribution of dislocation starvation to the total plastic strain is not investigated. Similarly, the nanopillar size limits below which the 
dislocation starvation occurs is not clearly established. In view of this, a systematic MD simulations have been performed on Cu nanopillars with size ranging 
from 5 to 21.5 nm to investigate the dislocation starvation mechanism under the compressive loading. The size dependence of dislocation starvation and its 
contribution to the total plastic strain have been investigated in detail.

\section{Simulation Details}

\begin{figure}[h]
\centering
\includegraphics[width=4.0cm]{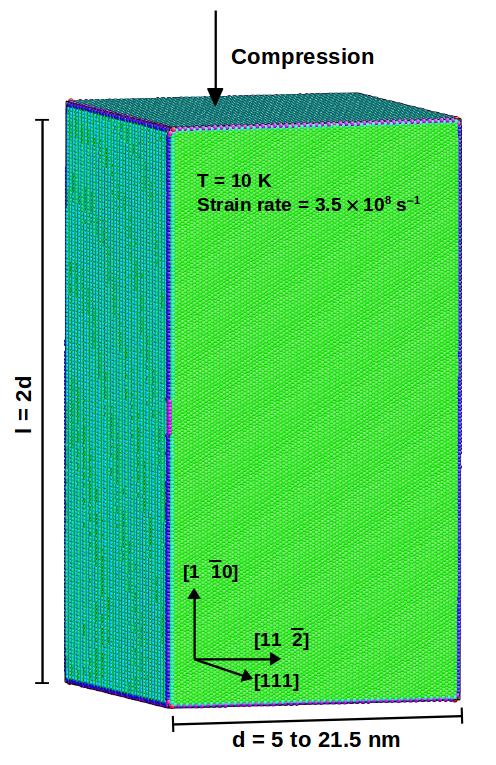}
\caption {The initial structure of the Cu nanopillar considered in this study. It contains the details of orientation, size, length or aspect ratio and loading 
direction. The atoms are coloured according to their potential energy.}
\label{Initial}
\end{figure}

All MD simulations have been carried out in Large scale Atomic/Molecular Massively Parallel Simulator (LAMMPS) package \cite{LAMMPS}. Embedded atom method 
(EAM) potential for Cu given by Mishin and co-workers \cite{potential} has been used for describing the inter-atomic forces between copper atoms. This EAM 
potential is widely used to study the plastic deformation related studies in Cu nanowires/nanopillars \cite{Liang-PRB,Sainath-PLA}. As shown in Figure 
\ref{Initial}, single crystal Cu nanopillars oriented in $[1\bar1 0]$ axial direction with (111) and $(11\bar2)$ as side surfaces were considered for this 
study. The $<$110$>$ orientation is selected mainly because of its wide and ease of synthesis \cite{JACS}. In the literature, it is widely synthesized 
orientation due to the preferential growth of Cu nanopillars along the $<$110$>$ direction, where the \{111\} planes are exposed to the surface, which have 
a lowest surface energy and hence stabilize the nanopillar \cite{JACS}. The model nanopillars had a square cross section shape with a cross section width (d) 
ranging from 5 to 21.5 nm (Figure \ref{Initial}). The pillar length (l) was twice the cross section width (d). In order to mimic the nanopillar surfaces, no 
periodic boundary conditions were used in any direction. Similar boundary conditions were used for nanopillars in earlier studies \cite{Sainath-PLA,Al-IJP,Twin-twin}. 
The model system was equilibrated to a required temperature of 10 K in NVT ensemble. Upon completion of equilibrium process, the compressive loading is applied 
in a displacement controlled manner at a constant strain rate of $3.5\times10^8$ s$^{-1}$ by imposing displacements to atoms along the loading axis ($[1\bar1 0]$ 
axis) that varied linearly from zero at the bottom layer to the maximum value at the top layer \cite{Sainath-PLA,Al-IJP,Twin-twin}. For all the nanopillars, 
the compressive loading has been performed up to a strain level of 0.25. The average stress is calculated from the Virial expression \cite{Virial}. In order 
to analyze the deformation, the visualization of atomic snapshots is accomplished using AtomEye \cite{Atomeye}. The Burgers vector of dislocations were 
determined by dislocation extraction algorithm (DXA) \cite{DXA} as implemented in OVITO \cite{OVITO} and were assigned according to Thomson tetrahedron.

\section{Results and Discussion}

\subsection{Stress-strain behaviour}

\begin{figure}
\centering

\begin{subfigure}[b]{0.425\textwidth}
\includegraphics[width=\textwidth]{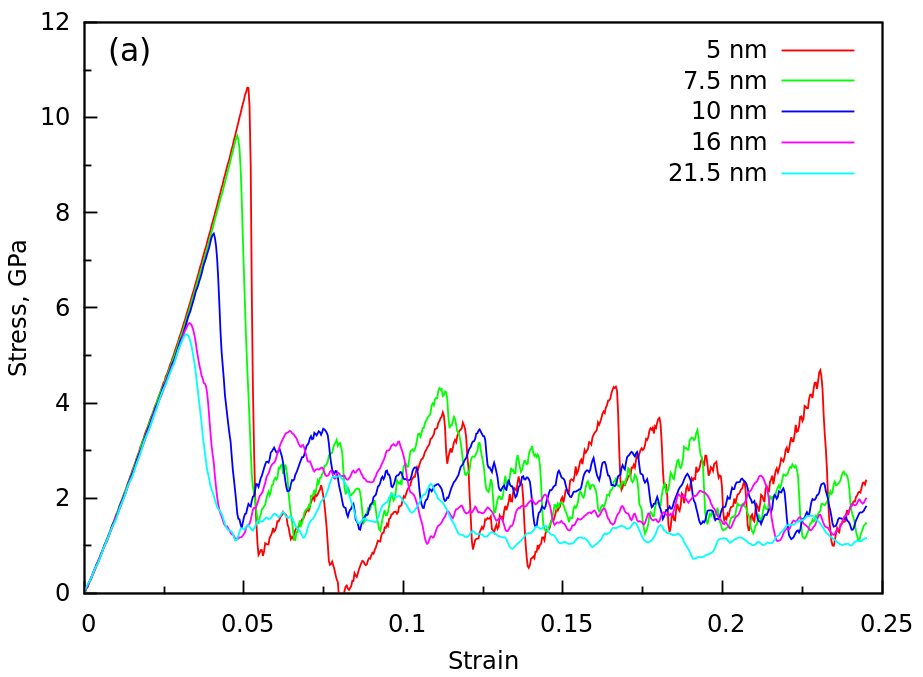}
\end{subfigure}
\qquad
\begin{subfigure}[b]{0.41\textwidth}
\includegraphics[width=\textwidth]{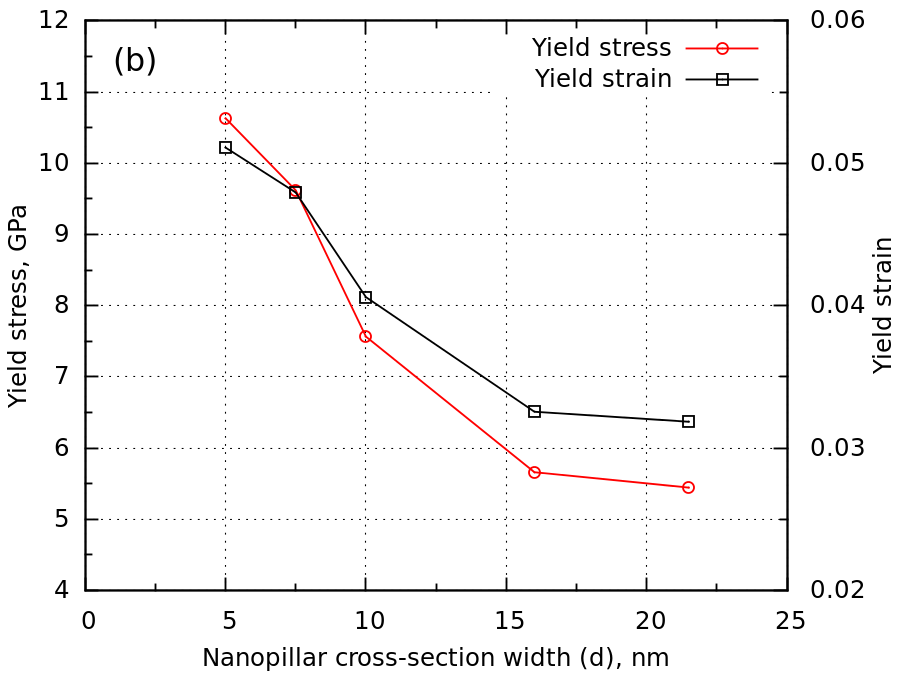}
\end{subfigure}
 \caption { (a) The stress-strain behaviour of $<$110$>$ oriented Cu nanopillars with cross-section width (d) ranging from 5 - 21.5 nm under compressive loading. 
 (b) The variation of yield stress and yield strain as a function of nanopillar size.} 
 \label{SS&YS}
 \end{figure}

The compressive stress-strain behaviour of $<$110$>$ oriented Cu nanopillars with cross-section width (d) ranging from 5 - 21.5 nm is presented in Figure 
\ref{SS&YS}a at 10 K. It can bee seen that all the nanopillars show similar stress-strain behaviour during the elastic deformation with an elastic modulus 
of 115 GPa. This value is in close agreement with experimental studies where the Young's modulus of $<$110$>$ Cu falls in the range 121-138 GPa \cite{YM-Cu}. 
Following the elastic deformation up to the peak, the flow stress drops abruptly to low level of stress indicating the occurrence of yielding in the 
nanopillars. Post yielding, large flow stress oscillations were observed for nanopillars with d = 5 to 10 nm, while these oscillations were minimal in 
magnitude for nanopillars with d = 16 and 21.5 nm. Moreover, most of these oscillations were of saw tooth type and self similar, i.e., similar to the 
initial stress-strain curve up to yielding. Figure \ref{SS&YS}a also suggest that the yield strength as well as the yield strain varies significantly with 
nanopillar size. The variation of compressive yield stress as well as yield strain as a function of nanopillar size is plotted in Figure \ref{SS&YS}b. It 
shows that with increasing size, the yield stress ($y_s$) and yield strain of the Cu nanopillars decreases rapidly for smaller nanopillars followed by marginal 
decrease at large sizes. The variation of yield stress as a function of nanopillar size $(d)$ obeys the power law relation $y_s = md^{-k}$, with m = 24.5 
and exponent (k) = 0.50. The observed values of the exponent(k) are in the range that has been reported experimentally for micron sized FCC pillars (0.3 - 1) 
\cite{exponent}.

 \subsection{Deformation mechanisms}

\begin{figure}
\centering
\includegraphics[width=6cm]{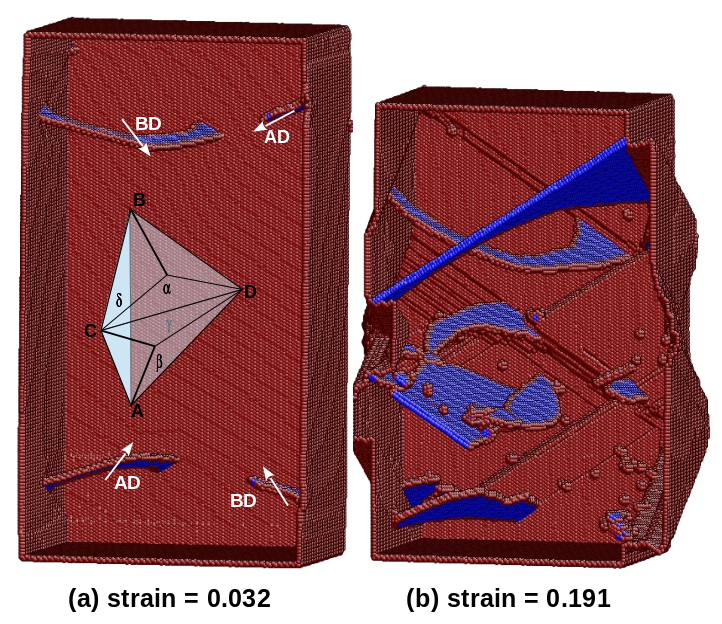}
\caption { The yielding (a), and subsequent plastic deformation behaviour (b), of $<$110$>$ oriented Cu nanopillars with d = 21.5 nm under compressive loading. 
The Thompson tetrahedron is shown in (a) to identify the slip systems. The atoms are colored according to the common neighbor analysis. The perfect fcc atoms 
are removed for clarity and atoms on stacking faults are shown in blue, while those of dislocations and surface are shown in red.}
\label{Yielding}
\end{figure}

In order to reveal the deformation mechanisms, the evolution of atomic configurations at various stages of deformation were analyzed for all the nanopillars. 
It has been observed that the deformation is dominated by the slip of extended dislocations and no twinning was noticed. Figure \ref{Yielding}a-b shows the 
typical yielding and subsequent plastic deformation behaviour of $<$110$>$ oriented Cu nanopillars with d = 21.5 nm under compressive loading. It can be seen 
that the nanopillar yields on two different slip planes, $\alpha$ and $\beta$ by the nucleation of a leading partial immediately followed by trailing partial 
on the same slip plane (Figure \ref{Yielding}a). The combination of leading and trailing partial dislocations separated by a stacking fault constitutes an 
extended dislocation, AD and BD. In an extended dislocation, a trailing partial erases the stacking fault produced by the glide of leading partial. With 
increasing strain, more and more such dislocations nucleate on different \{111\} planes (mainly $\alpha$ and $\beta$)  and interact with each other. The 
typical dislocation-dislocation interactions leading to the formation of many point defects are shown in Figure \ref{Yielding}b. The occurrence of full slip 
and the absence of twinning during the compressive loading of $<$110$>$ nanopillar can be understood by the Schmid factor analysis 
\cite{Park-JMPS,Weinberger-Cai,Rohith-CCM}. According to this analysis, the Schmid factor of trailing partial (0.471) is higher than the leading partial 
(0.235), which leads to easy nucleation of trailing partials in the nanopillar. Since the trailing partials cannot nucleate without the leading partials, 
the nucleation of leading partials automatically proceeds the trailing partials leading to an occurrence of full slip through two consecutive partial slips 
on the same \{111\} plane.

\subsection{Size effects and dislocation starvation}

\begin{figure}
\centering
\includegraphics[width=6cm]{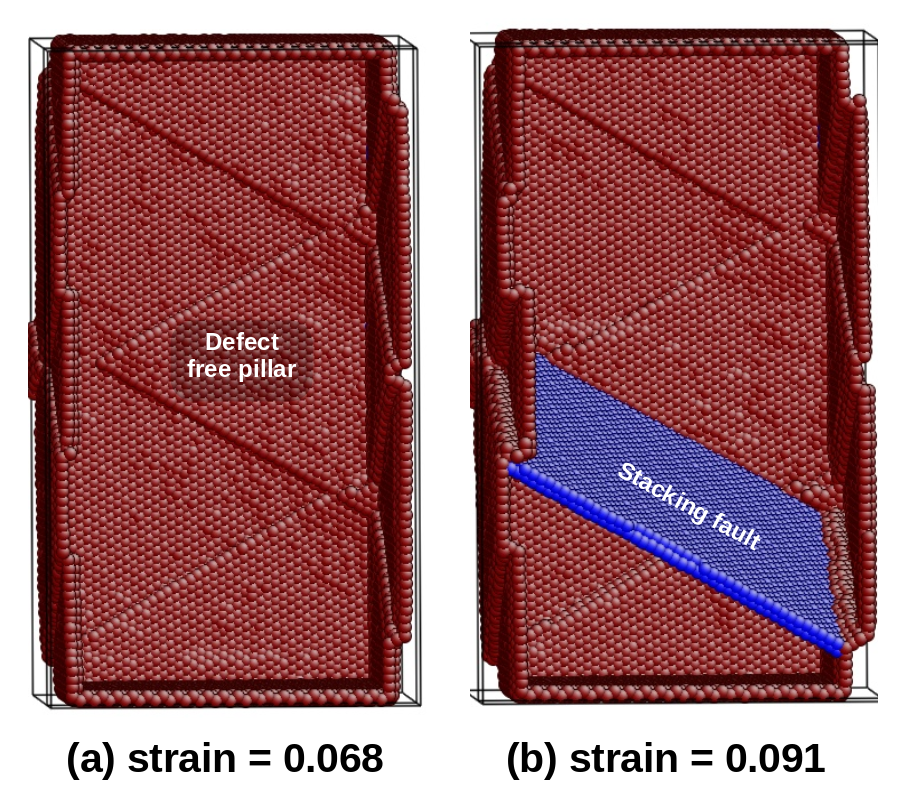}
\caption { The typical snapshots displaying the state of perfect dislocation starvation (a), and dislocation starvation with a wide stacking fault (b), during 
the plastic deformation of Cu nanopillar with d = 10 nm. The color coding of atoms is same as in Figure \ref{Yielding}.}
\label{Starvation}
\end{figure}

MD simulations on nanopillars with other sizes, i.e., d = 5, 7.5, 10 and 16 nm indicated that the yielding and the subsequent plastic deformation is dominated 
by the slip of extended dislocations. However, few important differences were observed during the plastic deformation of nanopillars with smaller sizes. 
Since the probability of dislocation interactions within the nanopillar decreases with decreasing size, the dislocations gliding on the slip plane easily 
escape to surface and leads to easy dislocation starvation in small size nanopillars. Figures \ref{Starvation}a-b show the typical snapshots displaying the 
state of perfect dislocation starvation and dislocation starvation with a wide stacking fault during the plastic deformation of nanopillar with d = 10 nm. 
Apart from starvation, many slip traces produced by the dislocations can also be seen on the surface of the nanopillar. \\

\begin{figure}
\centering
\includegraphics[width=7.5cm]{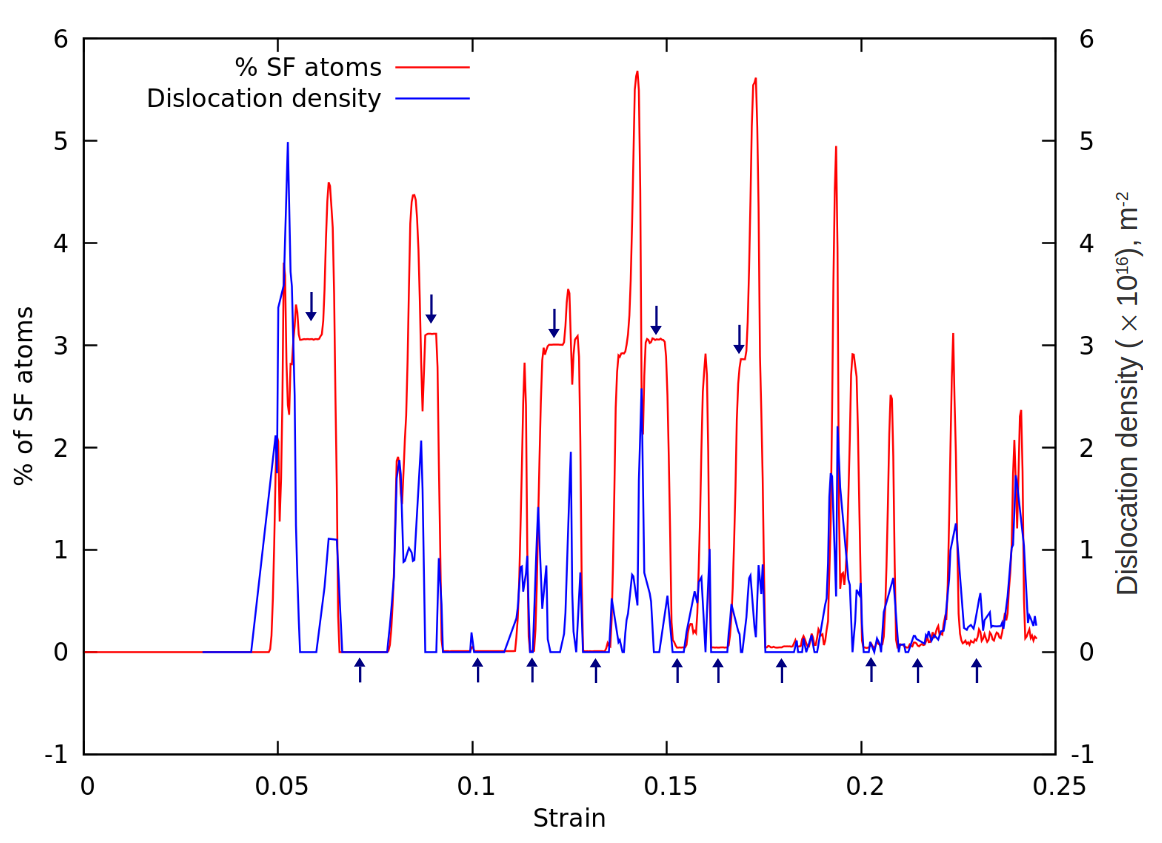}
\caption {The variation of percentage of stacking fault atoms along with dislocation density as a function of strain in the nanopillar with (d) = 7.5 nm. The 
short blue arrow marks indicates the dislocation starvation state, where the dislocation density is zero and the stacking fault atoms were either absent (zero) 
or constant.}
\label{DD-SF}
\end{figure}

\begin{figure}
\centering
\includegraphics[width=14cm]{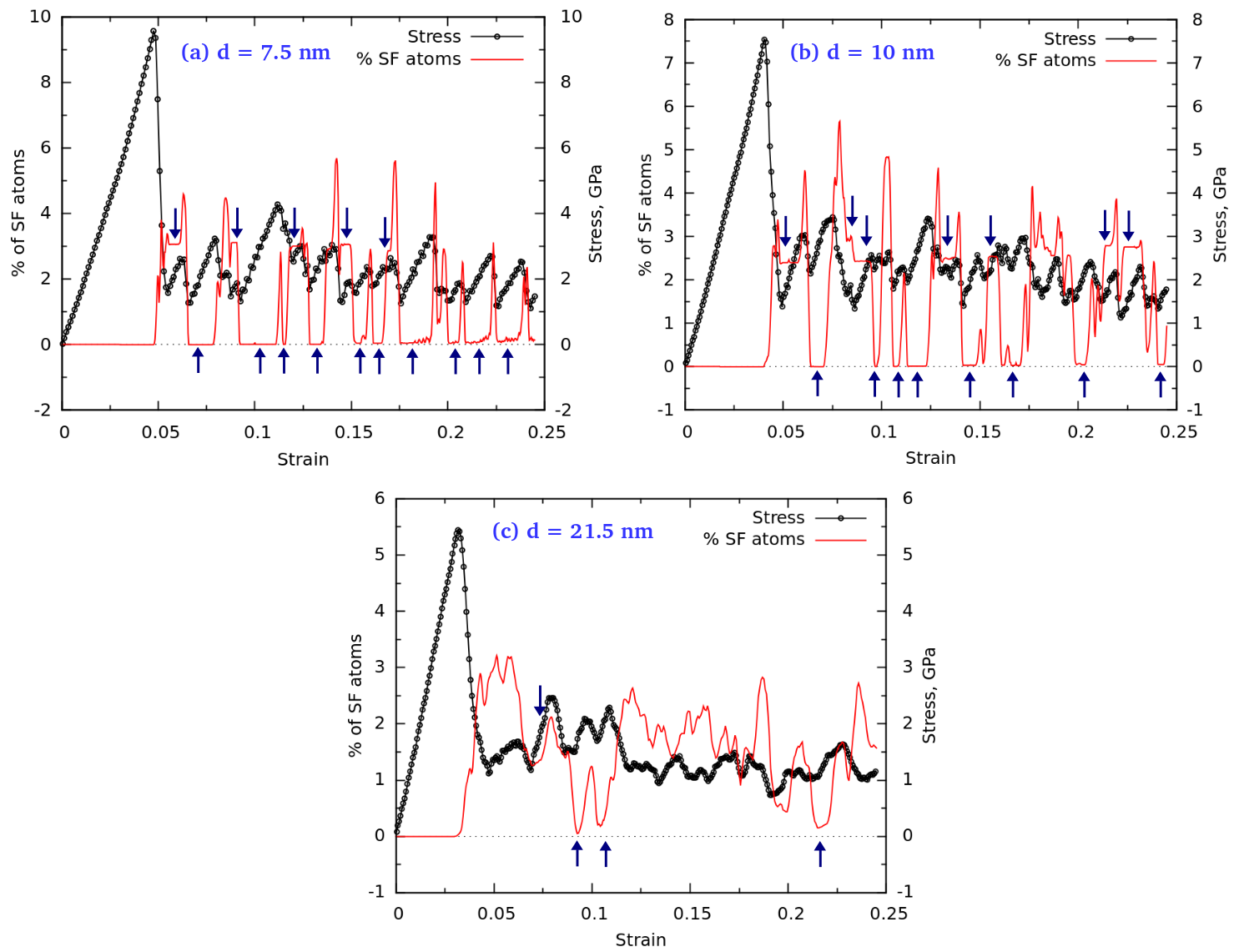}
\caption {The variation of percentage of stacking fault atoms along with flow stress as a function strain for the nanopillars with (a) d = 7.5 nm, (b) d = 10 nm, 
and (c) d = 21.5 nm. The short blue arrow marks indicates the dislocation starvation state.}
\label{SF-atoms}
\end{figure}

In order to find the frequency of occurrence of dislocation starvation and also its contribution to plastic strain, the number of stacking fault atoms were 
measured as a function of strain for all the nanopillars. Since the deformation is dominated by the extended dislocations (partial dislocations separated by 
stacking faults), the absence of stacking fault atoms and/or zero change in the stacking fault atoms over a period of strain indicates dislocation starvation 
in the nanopillar (Figure \ref{DD-SF}). It can be seen that whenever the dislocation density is zero, i.e., dislocation starvation, the stacking fault atoms 
were either absent (zero) or constant. This suggest that the variation in \% SF atoms can be used to identify the dislocation starvation events in the nanopillars. 
Unlike dislocation density, the calculation of \% SF atoms requires minimal data and computation time, thus avoiding a tedious post-processing. Figures 
\ref{SF-atoms}a-c show the percentage of stacking fault atoms along with flow stress as a function of strain for the nanopillars with d = 7.5, 10 and 21.5 
nm, respectively. The dislocation starvation state is identified with blue arrow marks. It can be seen that for the nanopillar with d = 7.5 nm, the state of 
dislocation starvation occurs more frequently and each state occurring over extended period of strain (Figure \ref{SF-atoms}a). With increasing the nanopillar 
size, the frequency of occurrence of dislocation starvation along with the duration of each state decreases (Figure \ref{SF-atoms}b) and finally negligible 
dislocation starvation is observed in the nanopillar with d = 21.5 nm (Figure \ref{SF-atoms}c). Further, it can be seen that when the nanopillar is in dislocation 
starvation state, the flow stress increases linearly indicating that the nanopillar undergoes repeated elastic deformations due to frequent dislocation 
starvation \cite{ACao}. As a result of dislocation starvation, the strength of the nanopillar increases significantly. 

\subsection{Dislocation starvation strain}
 
\begin{figure}[h]
\centering
\includegraphics[width=7cm]{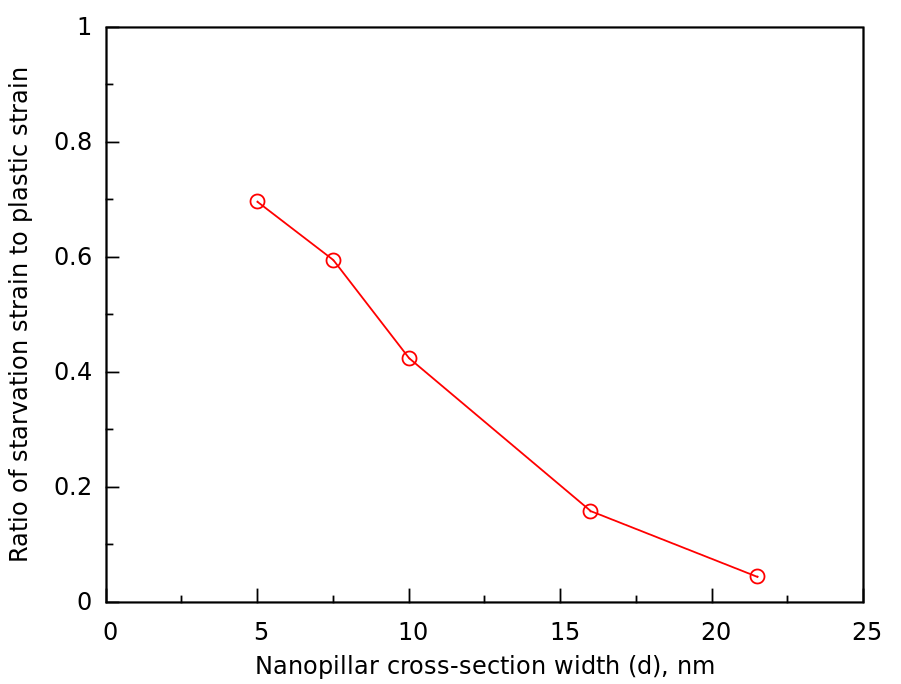}
\caption {The variation of the ratio of dislocation starvation strain to that of the plastic strain as a function of nanopillar size.}
\label{Starve-contrib}
\end{figure}

In order to characterize the contribution of dislocation starvation to the plastic strain or in turn plastic deformation, the aggregate strain over which 
the dislocation starvation occurs during the plastic deformation is measured for all the nanopillar sizes. This is obtained by combining the strain duration 
corresponding to the each arrow mark shown in Figure \ref{SF-atoms}. Henceforth, this strain is termed as dislocation starvation strain. The value of 
dislocation starvation strain varies from 0.14 in nanopillar with d = 5 nm to a negligible 0.01 in nanopillar with d = 21.5 nm. However, the dislocation 
starvation strain alone may not be a useful parameter. To obtain the contribution of dislocation starvation to the plastic strain, the ratio of dislocation 
starvation strain to that of the plastic strain is calculated as a function of nanopillar size and presented in Figure \ref{Starve-contrib}. It can be seen 
that the ratio decreases from 0.7 to 0.05, when the size increased from 5 nm to 21.5 nm. This indicates that for nanopillar with d = 5 nm, over 70\% of the 
duration of the plastic deformation is spent in dislocation starvation state. The duration of dislocation starvation state decreases with increasing size 
and reduces to below 5\% for the nanopillar with d = 21.5 nm (Figure \ref{Starve-contrib}). This suggest the existence of a threshold cross-section width 
of about 20 nm above which the dislocation starvation is almost negligible, ie., doesn't occur. Recently, in nanopillars of high entropy alloys, the critical 
size for the occurrence of dislocation starvation state is found to be around 15 nm and in perfect Cu nanopillars it is found to be close to 30 nm 
\cite{Starvation-HEA}. These results confirm that there exists a length scale in nanopillars below which the dislocation starvation can be observed. The 
main reasons for the observation of dislocation starvation in small size pillars is high image forces on dislocations coupled with less travel distance 
leading to easy escape of dislocations to the nearby free surfaces \cite{Greer-Nix,Greer-TSF}. On the other hand, the small image forces along with longer 
travel distance increase the probability of dislocation interaction and multiplication resulting in nanopillars with large size populated with dislocations 
all the time \cite{Greer-Nix,Greer-TSF}. Further, when the nanopillar is in dislocation starvation state, it deforms elastically even during the plastic 
deformation until new dislocations nucleate from the surfaces. This leads to the observation of saw tooth type large flow stress oscillations in small size 
nanopillars as shown in Figure \ref{SS&YS}a. Conversely, a negligible dislocation starvation in large size nanopillars caused minimal stress oscillations 
as seen from Figure \ref{SS&YS}a.

\subsection{Dislocation starvation and twinning}

\begin{figure}[h]
\centering
\includegraphics[width=5cm]{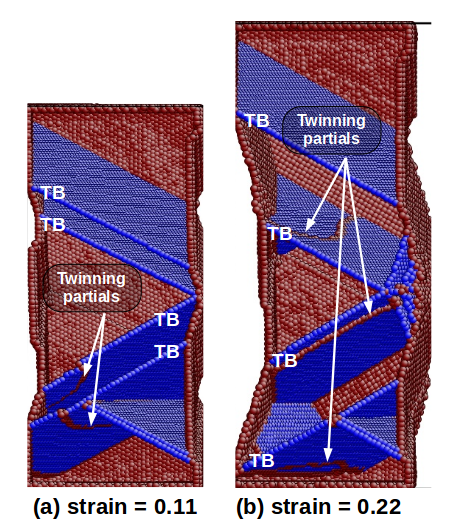}
\caption {The deformation behaviour of $<$110$>$ Cu nanopillars with d = 10 nm under tensile loading. The deformation proceeds by twinning mechanism and the 
twin boundaries are always populated with twinning partials during twin growth, i.e, plastic deformation stage.}
\label{Tension-deform}
\end{figure}

The present results indicate that the $<$110$>$ Cu nanopillars which deform by an extended dislocation slip exhibit an extensive dislocation starvation 
mechanism especially in small size pillars with d $<$ 20 nm. In this context, it is interesting to understand the possibilty of dislocation starvation 
when the nanopillar deforms by twinning mechanism. Since $<$110$>$ oriented FCC nanopillars are expected to deform by twinning under tensile loading 
\cite{Weinberger-Cai,Rohith-CCM}, additional MD simulations have been carried out on the tensile loading $<$110$>$ Cu nanopillars with size d = 10 nm. 
The results indicate that the deformation proceeds extensively by deformation twinning on two different twin systems (Figures \ref{Tension-deform}a-b). 
The twin grows by the repeated nucleation and glide of twinning partials on twin boundaries. Compared to trailing partial nucleation which is required 
for extended dislocation slip, the nucleation of twinning partials require low stress \cite{TBM,TBM-lowstress}. As a result, a new twinning partial 
nucleate before the annihilation of the existing one, i.e., there is a continuous supply of twinning partials. Due to this, the twin boundaries are 
always populated with twinning partials and the dislocation starvation is rarely observed in case of nanopillars deforming by twinning mechanism 
(Figure \ref{Tension-deform}). Due to the absence of dislocation starvation, the flow stress during the plastic deformation of nanopillar deforming 
by twinning shows no significant saw-tooth type fluctuations \cite{Sainath-PLA,Rohith-CCM} in contrast to that shown in Figure \ref{SS&YS}a. These 
results show that the $<$110$>$ Cu nanopillars exhibit tension-compression asymmetry in deformation mechanism as well as in dislocation starvation. 
The other important aspect of dislocation starvation is that it is observed mainly in FCC nanopillars. The dislocation starvation mechanism has not 
been observed in BCC nanopillars \cite{Cai-PNAS,Sai-MSEA}. The high mobility coupled with a planer core structure enables easy escape of the dislocations 
to surfaces leading to dislocation starvation state in FCC nanopillars. Whereas in BCC nanopillars, the dislocations self multiply themselves before 
their annihilation \cite{Cai-PNAS,Sai-JAP} as result of which, the nanopillars are always populated with dislocations without any starvation.


\subsection{Effect of temperature, strain rate and stacking fault energy}
Since the dislocation starvation is mainly due to the easy escape of dislocations to the surface, it is interesting to understand how the temperature, 
strain rate and stacking fault energy (SFE) influences the dislocation starvation mechanism in nanopillars. In nanoscale materials with high SFE, a pronounced 
dislocation starvation is reported \cite{Starvation-SFE}. The high SFE suppresses the dissociation of full dislocations into partials and hence results 
in easy escape of dislocations to the surface. Previous MD simulation study on $<$100$>$ Cu nanowires has shown that for a given size, the exhaustion rates of 
dislocations increases with increasing temperature and strain rate \cite{Rohith-Physica}. In other words, due to high velocity, the dislocations can 
easily escape to surface at high temperature and strain rates \cite{Temp-Disl-vel,Srate-Disl-vel}. As a result, the nanopillar can easily go to dislocation 
starvation state. In contrast to this observation in Cu nanopillars, the dislocation dynamics (DD) simulations study on BCC Nb micropillars has shown that the 
suppression of thermally activated cross-slip at low temperatures prevents dislocation multiplication and thus results in easy dislocation starvation 
\cite{Temp-starvation}. These contrasting observation in nanopillars and micropillars indicate that the effect of temperature on dislocation starvation 
in metallic nanopillars needs a detailed investigation. Apart from these factors, it will also be interesting to explore the effect of twin boundaries, which 
act as a strong obstacle to dislocation motion, on dislocation starvation.

\section{Conclusions}

Atomistic simulations have been performed to understand the size effects on the dislocation starvation mechanism in Cu nanopillars under compressive loading. 
Based on this study, the following conclusions have been drawn:

(1) The $<$110$>$ Cu nanopillars with size (d) ranging from 5 to 21.5 nm deform by the slip of an extended dislocations and exhibit dislocation starvation mainly 
in small size nanopillars with d $<$ 20 nm.

(2) The frequency of the occurrence of dislocation starvation state is highest in small size nanopillar with d = 5 nm and it decreases with increasing nanopillar 
size. Above 20 nm size, no incident of dislocation starvation has been observed.

(3) The dislocation starvation strain has been defined in this study and based on this, the contribution of dislocation starvation to the total plastic strain 
or plastic deformation is characterized. The contribution of dislocation starvation state to the total plastic strain is measured in terms of the ratio of 
dislocation starvation strain to that of the plastic strain. This ratio decreases from 0.7 to 0.05 when the size is varied from 5 nm to 21.5 nm. This indicates 
that for nanopillar with d = 5 nm, over 70\% of the duration of the plastic deformation is spent in dislocation starvation state. This duration of dislocation 
starvation state decreases with increasing size and reduces to below 5\% for the nanopillar with d = 21.5 nm. 

(4) The variation of yield stress ($y_s$) as a function of nanopillar size $(d)$ obeys the power law relation $y_s = md^{-k}$, with m = 24.5 and exponent (k) = 0.50.

(5) The dislocation starvation is not observed even in small size nanopillars when the deformation is dominated by twinning mechanism. This suggest that the 
dislocation starvation is a dominant phenomena when the deformation is dominated by the slip of partial/extended dislocations.

\section*{Disclosure statement}
The authors report that there are no competing interests to declare.

\section*{Funding}

This work received no funding.

\section*{Data availability}

The data that support the findings in this paper are available from the corresponding author on request.

\end{document}